\documentstyle[fleqn]{tp}

\input psfig

\def\sec{\ifmmode \,\, {\rm sec} \else sec \fi}
\def\eV {\ifmmode \,\, {\rm eV} \else eV \fi}
\def\keV{\ifmmode \,\, {\rm keV} \else keV \fi}
\def\MeV{\ifmmode \,\, {\rm MeV} \else MeV \fi}
\def\GeV{\ifmmode \,\, {\rm GeV} \else GeV \fi}
\def\TeV{\ifmmode \,\, {\rm TeV} \else TeV \fi}
\def\fm{\ifmmode \,\, {\rm fm} \else TeV \fi}
\def\pbarn{\ifmmode \,\, {\rm pb} \else pb \fi}
\def\km{\ifmmode {\rm km}\, \else km \fi}
\def\Mpc{\ifmmode {\rm Mpc}\, \else Mpc \fi}
\def\Gyr{\ifmmode {\rm Gyr}\, \else Gyr \fi}

\def\fun#1#2{\lower3.6pt\vbox{\baselineskip0pt\lineskip.9pt
  \ialign{$\mathsurround=0pt#1\hfil##\hfil$\crcr#2\crcr\sim\crcr}}}
\def\la{\mathrel{\mathpalette\fun <}}
\def\ga{\mathrel{\mathpalette\fun >}}
\def\order{{\cal O}}

\def\sbar#1{\kern 0.8pt
        \overline{\kern -0.8pt #1 \kern -0.8pt}
        \kern 0.8pt}  

\def\meter{\ifmmode \,\, {\rm m} \else m \fi}
\def\yr {\ifmmode \,\, {\rm yr} \else yr \fi}

\def\sr{\ifmmode \,\, {\rm sr} \else sr \fi}

\def\hatn{{\bf \hat n}}

\begin{document}

\twocolumn[
\title{The Cosmic Microwave Background: Beyond the Power Spectrum}
\author{Marc Kamionkowski \\
{\it Department of Physics, Columbia University} \\
{\it 538 West 120th Street, New York, NY 10025 U.S.A.} \\
{\tt kamion@phys.columbia.edu}}
\vspace*{16pt}   

ABSTRACT.\
Much recent work on the cosmic microwave background (CMB) has
focussed on the angular power spectrum of temperature
anisotropies and particularly on the recovery of cosmological
parameters from acoustic peaks in the power spectrum.  However,
there is more that can conceivably be done with CMB measurements.
Here I briefly survey a few such ideas:
cross-correlation with other cosmic backgrounds as a probe of
the density of the Universe; CMB polarization as a
gravitational-wave detector; secondary anisotropies and the
ionization history of the Universe; tests of alternative-gravity 
theories; polarization, the Sunyaev-Zeldovich effect, and cosmic 
variance; and tests for a neutrino mass.
\endabstract]

\markboth{Marc Kamionkowski}{The Cosmic Microwave Background:
Beyond the Power Spectrum}

\small

\section{Introduction}

Extraordinary results from COBE and balloon-borne and
ground-based experiments, a recent healthy interplay between
theorists and experimentalists, and the vision of MAP and Planck
on the horizon have focussed a considerable amount of attention
on the cosmic microwave background (CMB).  The primary aim of
these experiments is recovery of the angular power spectrum of
the temperature anisotropy, and from this, we hope to determine
cosmological parameters and test structure-formation theories.
However, there is conceivably much more that can be learned from 
the CMB.  Here I review briefly (and by no means exhaustively!)
a few such ideas.  Although the title suggests otherwise, some
of the ideas discussed {\it do} involve the power spectrum.

\section{Cross-Correlation with Other Backgrounds}

If the Universe has a critical density, then temperature
fluctuations are produced by density perturbations at or near
the surface of last scatter at a redshift $z\simeq1100$, well
beyond the reach of even the deepest galaxy surveys.  However,
if the Universe has less than the critical density, either in a
flat cosmological-constant or open Universe, then additional
anisotropies are produced by the red- and blue-shifting of
photons as they pass in and out of gravitational potential wells
along the line of sight at redshifts $z \la
\Omega_0^{-1}-1$.  Therefore, if $\Omega_0<1$, then there should 
be some cross-correlation between the CMB temperature in a given 
direction and some tracer of the mass distribution along that
same line of sight (Crittenden \& Turok 1996), such as the
extragalactic x-ray background.

\begin{figure*}
\centering\mbox{\psfig{figure=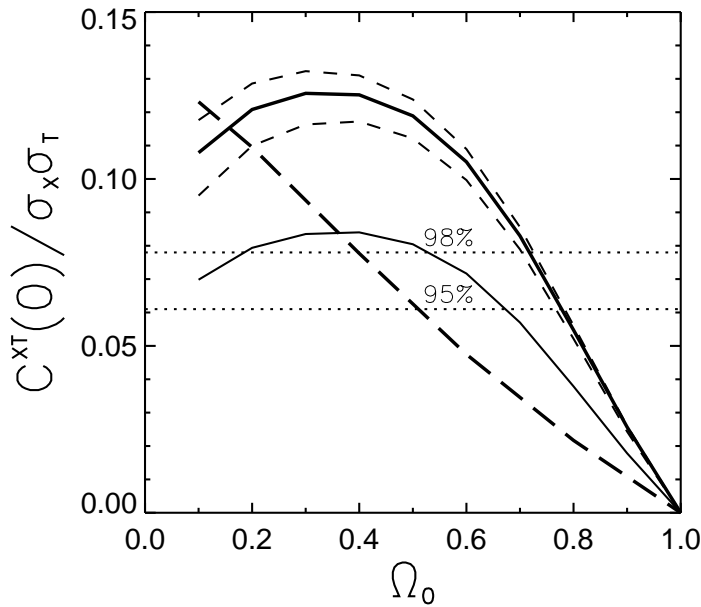,width=9.75cm}}
\caption[]{CMB/XRB cross-correlation amplitude in a
     flat cosmological-constant (heavy dashed curve), and open
     (heavy solid) curve for a CDM-like power spectrum with
     $\Gamma \simeq \Omega_0 h=0.25$, scaled
     by the CMB and XRB rms fluctuation amplitudes $\sigma_T$
     and $\sigma_X$.    The lighter dashed curves illustrate the
     uncertainty in the prediction due to uncertainties in the
     XRB redshift distribution.  The lighter solid curve is the
     open-Universe result for $\Gamma=0.5$ (but note that this
     requires an untenable value of $h>1$ for $\Omega_0<0.5$).
     The dotted curves show upper limits from
     Boughn, Crittenden \& Turok (1998).  [From
     Kinkhabwala \& Kamionkowski (1998).]}
\label{fig:ali}
\end{figure*}

Fig. \ref{fig:ali} shows predictions for the cross-correlation
amplitude in flat cosmological-constant and open models as a
function of $\Omega_0$, as well as experimental upper limits.
The results seem to indicate that an open universe must have
$\Omega_0\ga 0.7$, and that $\Omega_0\simeq0.3$ is ruled
out.  However, the theoretical predictions assume that all of
the XRB fluctuation is due to density perturbations; this
implies an x-ray bias $b_x \simeq 3$.  However, some of the XRB
fluctuation may simply be due to Poisson fluctuations in the
number of x-ray sources.  If so, the inferred x-ray bias is
lowered accordingly, and the predicted cross-correlation is also 
decreased by the same factor.  Since the scaled
cross-correlation amplitude is $\simeq0.13$ at
$\Omega_0\simeq0.3$, and the 95\% CL upper limit is
$\simeq0.6$, it suggests that an open-CDM Universe is viable
only if the x-ray bias is $b_x\la1.5$, significantly
smaller than biases of other high-redshift populations.

\section{Reionization and Secondary Anisotropies}
Although most of the matter in CDM models does not undergo
gravitational collapse until relatively late in the history of
the Universe, some small fraction of the mass is expected to
collapse at early times.  Ionizing radiation released by this
early generation of star and/or galaxy formation will partially
reionize the Universe, and these ionized electrons
will re-scatter at least some cosmic microwave background (CMB)
photons after recombination
at a redshift of $z\simeq1100$.  Theoretical
uncertainties in the process of star formation and the resulting
ionization make precise predictions of the ionization history
difficult.  Constraints to the shape of the CMB blackbody
spectrum and detection of CMB anisotropy at degree angular
scales suggest that if reionization occurred, the fraction of
CMB photons that re-scattered is small.  Still, estimates show
that even if small, at least some reionization is expected in
CDM models: for example, the most careful recent calculations
suggest a fraction $\tau_r\sim0.1$ of CMB photons were
re-scattered (Haiman \& Loeb 1997). 

Scattering of CMB photons from ionized
clouds will lead to anisotropies at arcminute separations below
the Silk-damping scale (the Ostriker-Vishniac
effect) (Ostriker \& Vishniac 1985; Vishniac 1987; Jaffe \&
Kamionkowski 1998).  These
anisotropies arise at higher order in perturbation theory and
are therefore not included in the usual Boltzmann calculations
of CMB anisotropy spectra.  The level of anisotropy is
expected to be small and it has so far eluded detection.  However,
these anisotropies may be observable with forthcoming CMB
interferometry experiments that probe the
CMB power spectrum at arcminute scales.

\begin{figure*}
\centering\mbox{\psfig{figure=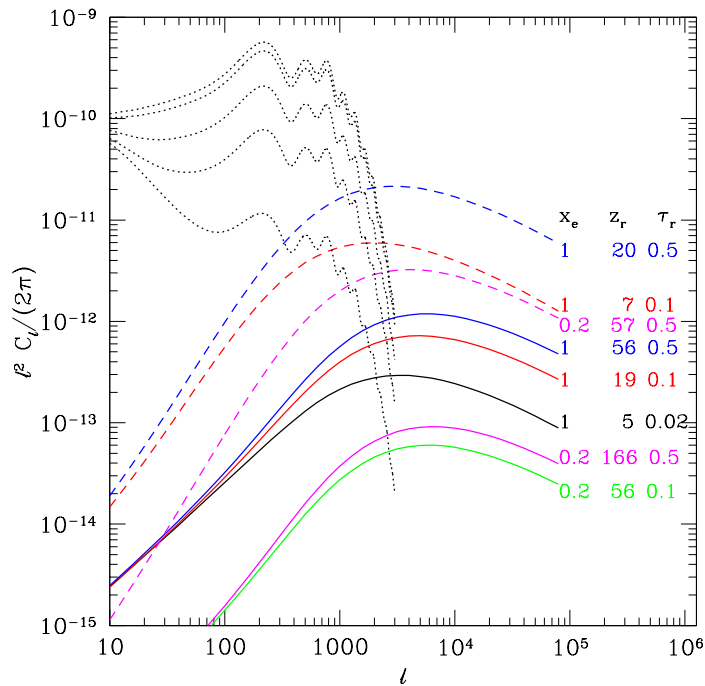,width=9.65cm}}
\caption[]{Multipole moments for the Ostriker-Vishniac effect for
     the COBE-normalized canonical standard-CDM model
     ($\Omega=1$, $h=0.5$, $n=1$, $\Omega_b h^2 =0.0125$), for a
     variety of ionization histories.as listed. 
     We also show predictions for several open
     high-baryon-density models with the same $x_e$ and
     $\tau_r$, normalized to the cluster abundance, with dashed
     curves.  The dotted curves show the primary anisotropy for
     this model for $\tau_r=0.0$, 0.1, 0.5, 1, and 2, from top
     to bottom.  [From Jaffe \& Kamionkowski (1998).]}
\label{fig:Clsplot}
\end{figure*}

Fig.~\ref{fig:Clsplot} shows the predicted
temperature-anisotropy spectrum from the Ostriker-Vishniac
effect for a number of ionization histories.  The
ionization histories are parameterized by an ionization fraction
$x_e$ and a redshift $z_r$ at which the Universe becomes
reionized.  The optical depth $\tau$ to the
standard-recombination surface of last scatter can be obtained
from these two parameters.  

Reionization damps the acoustic peaks in the
primary-anisotropy spectrum by $e^{-2\tau}$, as shown in
Fig.~\ref{fig:Clsplot}, but this damping is essentially
independent of the details of the ionization history.  That is,
any combination of $x_e$ and $z_r$ that gives the same $\tau$
has the same effect on the primary anisotropies.  So although
MAP and Planck will be able to determine $\tau$ from this
damping, they will not constrain the epoch of reionization.  On
the other hand, the secondary anisotropies (the
Ostriker-Vishniac anisotropies) produced at smaller angular
scales in reionized models {\it do} depend on the ionization
history.  For example, although the top and bottom dashed curves
in Fig.~\ref{fig:Clsplot} both have the same optical depth, they
have different reionization redshifts ($z_r=20$ and $z_r=57$).
Therefore, if MAP and Planck determine $\tau$, the amplitude of
the Ostriker-Vishniac anisotropy determines the reionization
epoch.

In a flat Universe, CDM models  normalized to cluster abundances
produce rms temperature anisotropies of 0.8--2.4 $\mu$K on
arcminute angular scales for a constant ionization fraction of
unity, whereas an ionization fraction of 0.2 yields rms
anisotropies of 0.3--0.8 $\mu$K.  In an open and/or
high-baryon-density Universe, the level of anisotropy is
somewhat higher.  The signal in some of these models may be
detectable with planned interferometry
experiments.

\section{Polarization and Gravitational Waves}

Although a CMB temperature map cannot unambiguously distinguish
between the density-perturbation and gravitational-wave
contributions to the large-angle CMB anisotropy, the two can be
decomposed in a model-independent fashion with a map of the CMB
polarization (Kamionkowski, Kosowsky \& Stebbins 1997a, 1997b; Seljak
\& Zaldarriaga 1997; Zaldarriaga \& Seljak 1997).  Suppose we
measure the linear-polarization ``vector'' $\vec P(\hatn)$ at
every point $\hatn$ on the sky.  Such a vector field can be
written as the gradient of a scalar function $A$ plus the curl
of a vector field $\vec B$,
\begin{equation}
     \vec P(\hatn) \, = \, \vec \nabla A \, + \, \vec\nabla \times \vec
     B.
\label{eq:curl}
\end{equation}
The gradient (i.e., curl-free) and curl components can be
decomposed by taking the divergence or curl of $\vec
P(\hatn)$ respectively.  Density perturbations are scalar metric
perturbations, so they have no handedness.  They can therefore
produce no curl.  On the other hand, gravitational waves {\it
do} have a handedness so they can (and we have shown that they
do) produce a curl.  This therefore provides a way to detect the
inflationary stochastic gravity-wave background and thereby
test the relations between the inflationary observables.  It
should also allow one to determine (or at least constrain in the
case of a nondetection) the height of the inflaton potential.

The sensitivity of a polarization map
to gravity waves will be determined by the
instrumental noise and fraction of sky covered, and by the
angular resolution.  Suppose the detector sensitivity is $s$ and
the experiment lasts for $t_{\rm yr}$ years with an angular
resolution better than $1^\circ$.  Suppose further that we
consider only the curl component of the polarization in our
analysis.  Then the smallest tensor amplitude ${\cal T}_{\rm
min}$ to which the experiment will be sensitive at $1\sigma$, in 
units of the measured COBE temperature quadrupole moment
$C_2^{\rm TT}$, is
(Kamionkowski \& Kosowsky 1998)
\begin{equation}
     {{\cal T}_{\rm min}\over 6\, C_2^{\rm TT}}
      \simeq 5\times 10^{-4} \left( {s\over \mu{\rm K}\,\sqrt{\rm
      sec}} \right)^2 t_{\rm yr}^{-1}.
\label{CCresult}
\end{equation}
Thus, the curl component of a full-sky polarization map is
sensitive to inflaton potentials $(V/m_{\rm Pl}^4)\ga 5 \times
10^{-15}t_{\rm yr}^{-1}$ $(s/\mu{\rm K}\, \sqrt{\rm sec})^2$.  
Improvement on current constraints with only the curl
polarization component requires a detector sensitivity
$s\la40\,t_{\rm yr}^{1/2}\,\mu$K$\sqrt{\rm sec}$.  For
comparison, the detector sensitivity of MAP will be $s={\cal
O}(100\,\mu$K$\sqrt{\rm sec})$.  However, Planck may conceivably
get sensitivities around $s=25\,\mu$K$\sqrt{\rm sec}$.  

A more complete analysis of the data will improve the
sensitivity relative to the simple estimate given
here. Furthermore, even a small amount of reionization will
actually increase the large-angle polarization (Zaldarriaga
1997) and thus also improve the detectability.

\section{New Particle and Gravitational Physics} 

The CMB may be used as a probe of new particle physics in yet
another way:  One of the primary goals of experimental particle
physics these days is pursuit of a nonzero neutrino mass.  Some
recent (still controversial) experimental results suggest that
one of the neutrinos may have a mass of $\order(5\,{\rm eV})$
(Athanassopoulos et al. 1995), and there have been some (again, still
controversial) arguments that such a neutrino mass is just what
is required to explain some apparent discrepancies between
large-scale-structure observations and the simplest
inflation-inspired standard-CDM model (e.g., Primack et al. 1995).

If the neutrino does indeed have a mass of $\order(5 \, {\rm
eV})$, then roughly 30\% of the mass in the Universe will be in
the form of light neutrinos.  These neutrinos will affect the
growth of gravitational-potential wells near the epoch of last
scatter, and they will thus leave an imprint on the CMB angular
power spectrum (Dodelson, Gates \& Stebbins 1996).  The effect
of a light neutrino
on the power spectrum is small, so other cosmological parameters
that might affect the shape of the power spectrum at larger
$l$'s must be known well.  Still, Hu, Eisenstein \& Tegmark (1998) argue
that by combining measurements of the CMB power spectrum with
those of the mass power spectrum measured by, e.g., the Sloan
Digital Sky Survey, a neutrino mass of $\order(5 \, {\rm eV})$
can be determined. 

The CMB may also conceivably be used to test alternative gravity 
theories (Liddle, Mazumdar \& Barrow 1998; Chen \&
Kamionkowski 1998) such as Jordan-Brans-Dicke or more
general scalar-tensor theories.  The idea here is that in such a 
theory, the expansion rate at the epoch of last scatter will be
different, and this will provide a unique signature in the CMB
power spectrum.

\section{Polarization in the SZ Effect}

Finally, there is a tremendous amount which can be learned from
scattering of CMB photons from the hot gas in x-ray clusters
both about cosmology and about cluster physics, much of which
has recently been reviewed by Birkinshaw (1998).  Most work on
the SZ effect to date has focussed on the temperature
perturbation produced.  However, several mechanisms may cause
these scattered photons to be polarized.  In particular, if the
radiation incident on a given cluster is anisotropic, the
scattered light will be polarized in proportion to the
quadrupole moment incident on this cluster.  In this way,
measurement of the SZ polarization may some day tell us about
the quadrupole anisotropy of the CMB as viewed by an observer in 
a distant cluster, and this could, in some sense, help reduce
the cosmic variance in the CMB determination of the large-scale
fluctuation amplitude (Kamionkowski \& Loeb 1997).

\section*{Acknowledgments}

I thank X. Chen, A. Jaffe, A. Kinkhabwala, A. Kosowsky, A. Loeb,
and A. Stebbins for collaboration on the projects described
here, and L. Wang for useful comments on this contribution.  This
work was supported by the U.S. D.O.E. Outstanding Junior
Investigator Award under contract DEFG02-92-ER 40699, NASA
NAG5-3091, and the Alfred P. Sloan Foundation.

\end{document}